\documentclass[a4paper,prx,aps,twocolumn,preprintnumbers,amsmath,amssymb,superscriptaddress]{revtex4-2}

\usepackage[utf8]{inputenc}
\usepackage{graphicx}
\usepackage{times}
\usepackage{amsmath}
\usepackage{amsfonts}
\usepackage{amssymb}
\usepackage{siunitx}

\usepackage{color}
\definecolor{red}{rgb}{0.8,0,0.1}

\definecolor{darkblue}{rgb}{0,0,.6}

\definecolor{lightgrey}{rgb}{0.7,0.7,0.7}

\definecolor{grey}{rgb}{0.4,0.4,0.4}

\begin{document}

\title{Open-cavity in closed-cycle cryostat as a quantum optics platform}

\author{Samarth Vadia}
    \email{samarth.vadia@physik.uni-muenchen.de}
\def\LMU{Fakult\"at f\"ur Physik, Munich Quantum Center,
  and Center for NanoScience (CeNS),
  Ludwig-Maximilians-Universit\"at M\"unchen,
  Geschwister-Scholl-Platz 1, 80539 M\"unchen, Germany}
\affiliation{\LMU}
\def\attocube{attocube systems AG,
  Eglfinger Weg 2, 85540 Haar bei M\"unchen, Germany}
\affiliation{\attocube}
\def\MCQST{Munich Center for Quantum Science and Technology (MCQST), Schellingtr. 4, 80799 M\"unchen, Germany}
\affiliation{\MCQST}

\author{Johannes Scherzer}
\affiliation{\LMU}

\author{Holger Thierschmann}
\affiliation{\attocube}

\author{Clemens Sch\"afermeier}
\affiliation{\attocube}

\author{Claudio Dal Savio}
\affiliation{\attocube}

\author{Takashi Taniguchi}
\affiliation{International Center for Materials Nanoarchitectonics, National Institute for Materials Science, 1-1 Namiki, Tsukuba 305-0044, Japan}

\author{Kenji Watanabe}
\affiliation{Research Center for Functional Materials, National Institute for Materials Science, 1-1 Namiki, Tsukuba 305-0044, Japan}

\author{David Hunger}
\affiliation{Karlsruher Institut f\"ur Technologie, Physikalisches Institut,
Institut f\"ur Quanten Materialien und Technologien, Wolfgang-Gaede-Str. 1, 76131 Karlsruhe, Germany}

\author{Khaled Karra\"i}
    \email{khaled.karrai@attocube.com}
\affiliation{\attocube}

\author{Alexander H\"ogele}
    \email{alexander.hoegele@lmu.de}
\affiliation{\LMU}
\affiliation{\MCQST}

\date{\today}

\begin{abstract}
The introduction of an optical resonator can enable efficient and precise interaction between a photon and a solid-state emitter. It facilitates the study of strong light-matter interaction, polaritonic physics and presents a powerful interface for quantum communication and computing. A pivotal aspect in the progress of light-matter interaction with solid-state systems is the challenge of combining the requirements of cryogenic temperature and high mechanical stability against vibrations while maintaining sufficient degrees of freedom for in-situ tunability. Here, we present a fiber-based open Fabry-P\'{e}rot cavity in a closed-cycle cryostat exhibiting ultra-high mechanical stability while providing wide-range tunability in all three spatial directions. We characterize the setup and demonstrate the operation with the root-mean-square cavity length fluctuation of less than $90$~pm at temperature of $6.5$~K and integration bandwidth of $100$~kHz. Finally, we benchmark the cavity performance by demonstrating the strong-coupling formation of exciton-polaritons in monolayer WSe$_2$ with a cooperativity of $1.6$. This set of results manifests the open-cavity in a closed-cycle cryostat as a versatile and powerful platform for low-temperature cavity QED experiments.
\end{abstract}
\maketitle

\section{INTRODUCTION}
The utilization of quantum physics promises to improve or even disrupt several technological fields such as computing, communication, simulation and metrology \cite{Deutsch2020}. The realm of systems with quantum applications covers a wide range of solid-state materials including quantum dots \cite{Imamoglu1999, Biolatti2000, Michler2017}, transition-metal dichalcogenides (TMDs) \cite{Wang2018, Mak2016}, color centers \cite{Hanson2008, Degen2017, Bradac2019} and rare-earth ions \cite{Thiel2011, Kindem2020}. These solid-state systems provide a light-matter interface for quantum optical protocols in a robust crystal matrix with full potential of integrability and scalability for future devices.

Controlling the interaction of matter with photons is essential for manipulating qubits to perform computational tasks as well as performing operations to transfer information from photons to solid-state emitters or vice versa. Photons are excellent as information carrier, however, it is extremely challenging to steer their interaction in a coherent manner. Therefore, a photonic resonator is an essential tool for establishing deterministic control over interactions in experiments of cavity quantum electrodynamics (cavity QED) \cite{Imamoglu1999, Zheng2000, Chang2014}. A promising platform for cavity QED is an open Fabry-P\'{e}rot cavity where a fiber-based micromirror forms a cavity with high finesse and minimal mode volumes with a macroscopic counterpart mirror that supports the solid-state system of interest \cite{Steinmetz2006, Hunger2010, Muller2010}. The advantages of such an open-cavity are two-fold. First, the open nature allows  easy integration of solid-state emitters inside the cavity. Second, it allows one mirror of the cavity to be moved with respect to the other for in-situ spectral tuning of the cavity resonance as well as lateral displacement which is advantageous to select the most suitable emitter region on the macroscopic mirror.

Solid-state emitters typically require operation at cryogenic temperature to reduce environmental decoherence effects. In view of the increasingly limited helium resources, it is expected that emerging quantum technologies will have to be compatible with closed-cycle cryostats which greatly facilitate stable and robust low-temperature operation \cite{Zhao2019}. However, the operation of a closed-cycle cryostat leads to mechanical vibrations that pose a significant challenge for a stable operation of open cavities. The mechanical stability required for an open-cavity of a given finesse $\mathcal{F}$ (or quality factor $Q=q\mathcal{F}$ of the cavity mode $q$) follows from the spatial-equivalent linewidth $\Delta L$ of the cavity transmission resonance at wavelength $\lambda$ as $\Delta L = \lambda/(2\mathcal{F})$. To ensure mirror-loss limited finesse, this resonance linewidth should not be compromised by mechanical vibrations leading to fluctuations in the cavity length. Thus, for a cavity of finesse $10^3$ (or Q-factor of the lowest cavity mode $q=1$) operated at $780$~nm wavelength, the mechanical stability should be better than $390$~pm, whereas deep sub-linewidth limit of $\Delta L / 10$ requires length fluctuations of less than $4$~pm for a finesse or lowest-mode Q-factor of $10^4$. In comparison, the mechanical vibrations induced by a typical closed-cycle cooler are in the order of \SI{10}{\micro\meter}. 

A useful open-cavity cryogenic platform has to address this primary challenge of achieving robust mechanical stability. Moreover, it should maintain all degrees of tunability and control for in-situ positioning of the two mirrors in all three spatial dimensions. The final key criterion of performance is the temperature as many solid-state emitters require operation below $10$~K. While substantial progress has been achieved in open Fabry-P\'{e}rot cavity experiments, in most cases, a tradeoff was made regarding at least one of the above criteria. There have been demonstrations of large tunability and high mechanical stability at room-temperature \cite{Barbour2011, Albrecht2013, Mader2015} or in experiments with liquid helium bath cryostat \cite{Besga2015, Riedel2017, Foerg2019, Najer2019, Tomm2021}. More recent experiments in a closed-cycle cryostat so far have achieved only limited mechanical stability, even in highly customized systems, and do not include or quantify in-situ large range tunability in various degrees of freedom \cite{Bogdanovic2017, Merkel2020, Salz2020, Casabone2020, Ruf2021}.

In this work, we present a fully-tunable open Fabry-P\'{e}rot cavity in a closed-cycle cryostat with a large range of lateral displacement. The integrated cavity reaches a temperature below $7$~K and combines displacement in three spatial directions over several mm with a free-space access to the cavity through an aspheric lens that can be displaced over mm, where all coarse tuning capabilities are provided with off-the-shelf nanopositioners. Additionally, we show that with the combination of passive and active vibration reduction techniques, the mechanical stability of the root-mean-square (rms) displacement of less than $90$~pm is achieved at the integration bandwidth of $100$~kHz. Subsequently, we demonstrate coherent interaction between excitons in a TMD with cavity photons. Using our tunable cavity platform with WSe$_2$ monolayer, we observe exciton-polaritons by controlling the energy detuning between cavity photons and TMD excitons.

\begin{figure*}[t]
\centering
\includegraphics[scale=1]{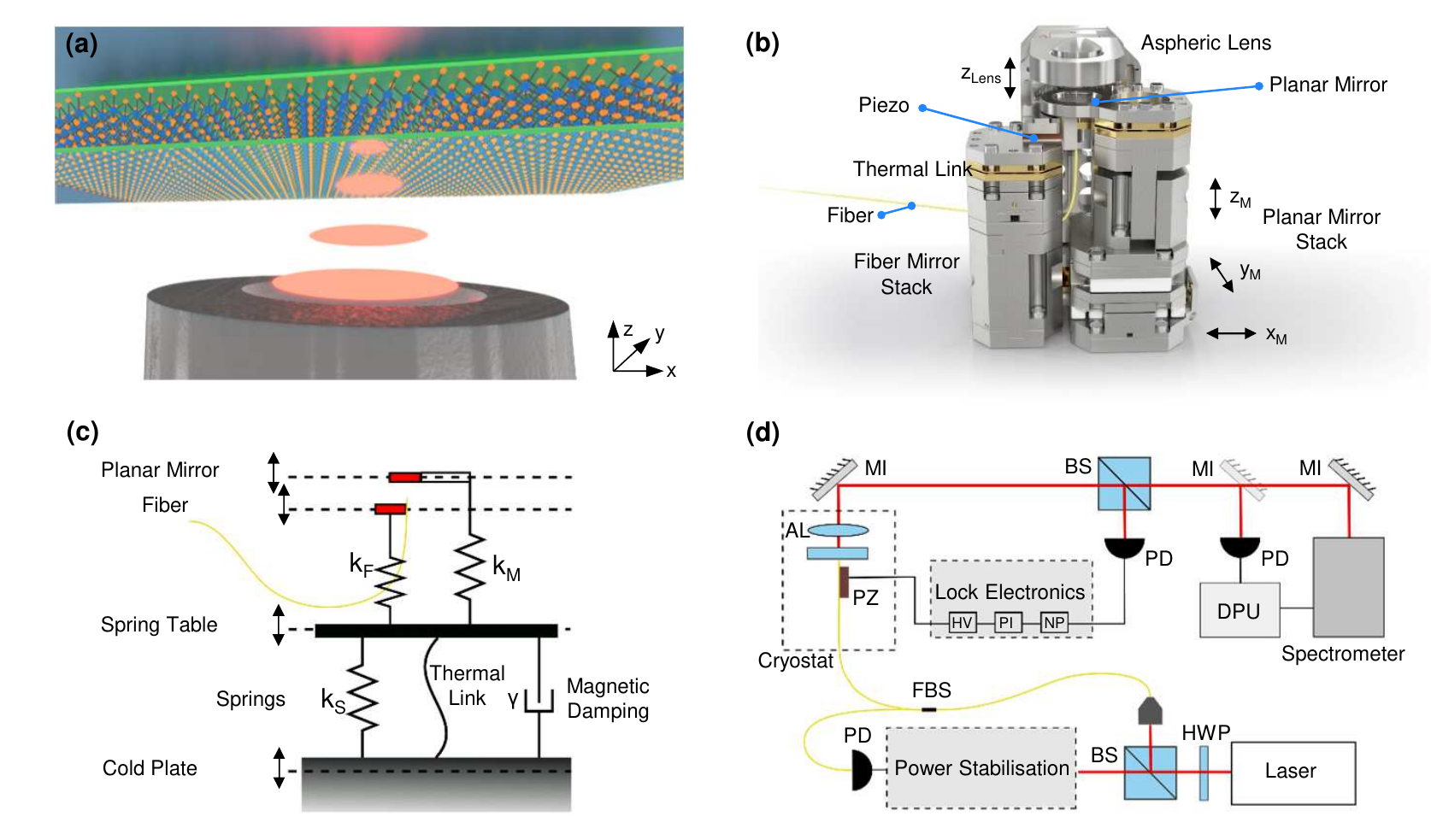}
\caption{Concept of an open-cavity in a closed-cycle cryostat. (a) An illustration of the cavity composed of a concave-profiled fiber mirror and a macroscopic mirror with monolayer WSe$_2$. (b) Sketch of the tunable cavity assembly that mounts on top of a vibration isolation stage. The stack on the right holds the planar mirror and consists of xyz-nanopositioner with a thermal link connecting the mirror to the cold plate to thermalize it to cryogenic temperature. The other stack on the left holds the fiber mirror and consists of a piezo element, thermal link and metal blocks to reach the same height as the right stack. The aspheric lens is mounted via another metal piece at the back with an additional nanopositioner to adjust the focal spot in z direction. (c) Schematic of the cavity assembly and vibration isolation stage built on top of the cold plate of the closed-cycle cryostat. Vibration isolation stage consists of the spring table placed on top of four springs (k$_{S}$) along with magnetic damping (damping constant $\gamma$), which is all placed on the cold plate. The thermal link enables cavity operation at cryogenic temperature. The cavity setup is represented as two springs on the spring table: planar mirror stack (with spring constant k$ _{M}$) and fiber mirror stack (with spring constant k$_{F}$). (d) Sketch of the experimental setup used for the stability characterization and strong-coupling operation. Assignment of abbreviations: $ \lambda $/2-waveplate (HWP), beamsplitter (BS), fiber beamsplitter (FBS), photodiode (PD), aspheric lens (AL), piezo element (PZ), mirror (MI), data processing unit (DPU); lock electronics: high voltage amplifier (HV), proportional-integral (PI) control electronics, notch pass (NP).}
\label{fig1}
\end{figure*}

\section{EXPERIMENTAL CRYO-CAVITY PLATFORM} \label{Sec2}
The design of our open Fabry-P\'{e}rot cavity is based on the combination of a fiber-micromirror and a macroscopic mirror. The micromirror is formed by a fiber end facet with a dimple that has been produced using CO$_2$ laser ablation technique \citep{Hunger2010}. Its macroscopic mirror counterpart supports a solid-state system of interest, as illustrated in Fig.~\ref{fig1}(a) for the specific case of semiconducting monolayer WSe$_2$. For the experiments reported here, both mirrors were coated with either a thin film of silver or a distributed Bragg reflector coating. 

The fiber and the macroscopic mirror were mounted in a configuration as shown in Fig.~\ref{fig1}(b). To ensure large tunability of the cavity not only along its optical axis (z) but also along lateral dimensions (x, y), we placed the two cavity mirrors on two separate mounts. The macroscopic mirror was fixed on top of a commercial xyz-nanopositioner (attocube, two ANPx311 and ANPz102) for precise and independent position control in the x and y direction over a range of $6$~mm each and cavity length control along z over a range of $4.8$~mm. The fiber mirror is mounted on another stack, consisting of rigid titanium blocks and a piezo, which only provides a tunability in the z direction through the piezo element for cavity fine tuning and active feedback control. The fiber position in the xy-plane is fixed. This ensures alignment of the optical axis of the cavity with an aspheric lens (Thorlabs, AL1210, NA = 0.55) mounted on a nanopositioner (attocube, ANPx311HL) with z displacement providing a free-space optical access to the cavity. 

\begin{figure}[t]
\centering
\includegraphics[scale=0.96]{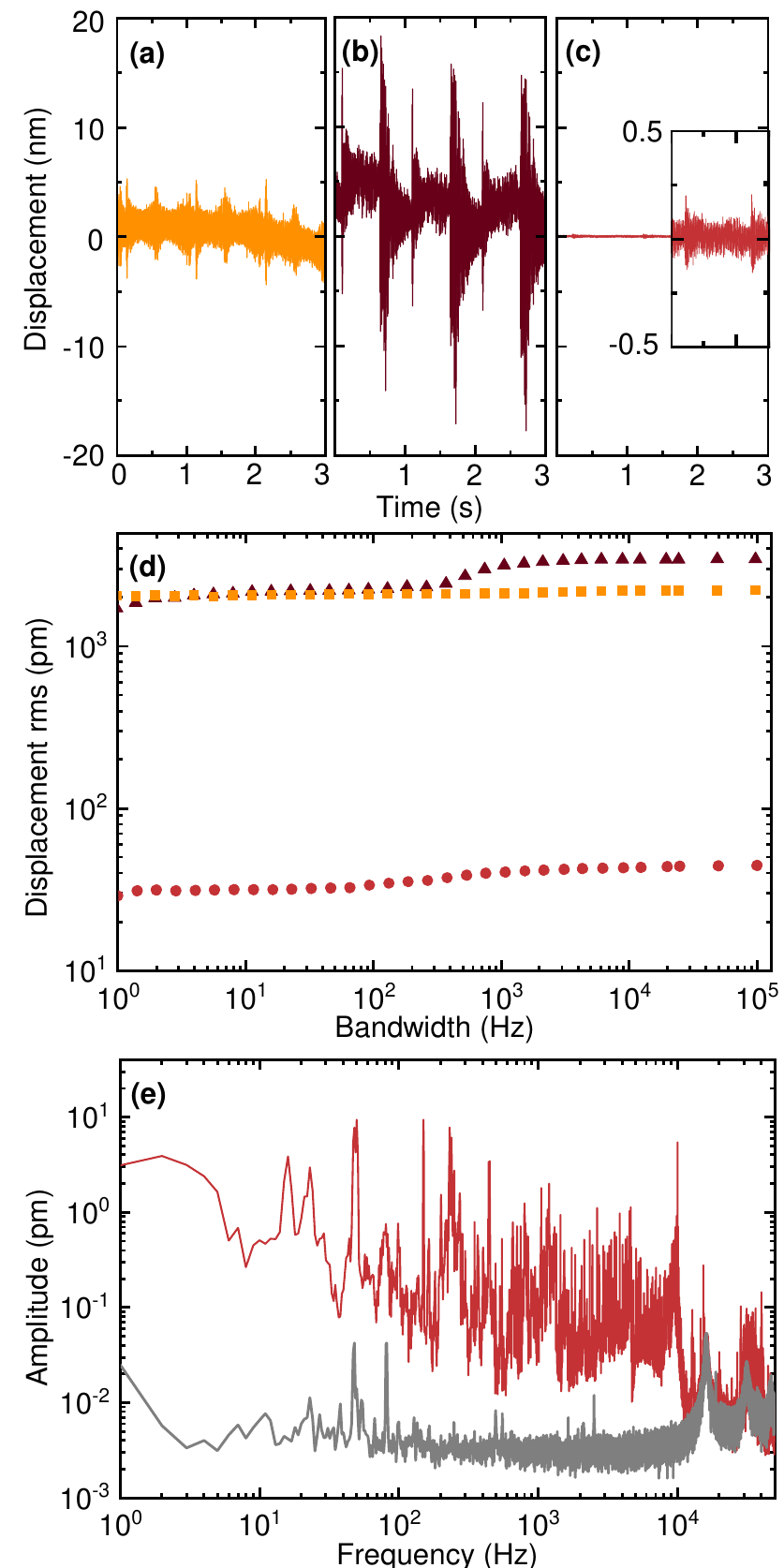}
\caption{Mechanical vibration characteristics at room-temperature. (a) Displacement of the cold plate of the cryostat. (b) Displacement of a xyz-nanopositioner stack on the cold plate. (c) Cavity length fluctuations of the open-cavity. All time trace data are shown for an integration bandwidth of $100$~kHz. (d) Displacement rms as a function of increasing measurement bandwidth for the cold plate (square), nanopositioner stack (triangle) and open-cavity (circle). (e) Fourier transform of cavity length fluctuations (red) and noise floor (grey) with a frequency resolution of $1$~Hz.}
\label{fig2} 
\end{figure}

The key challenge when using the above setup in a closed-cycle cryostat lies in the mechanical vibrations induced by the cooling cycle of the cryostat. The mechanical vibration amplitude at the cold head is in the range of 10 -- \SI{20}{\micro\meter}, mostly due to the impact of high pressure helium flow during the cooling cycle \cite{Tomaru2004}. The displacement amplitude is brought down to the level of a few nanometers at the cold plate of the cryostat (attocube, attoDRY800) that is aligned with the surface of the optical table \cite{Haft2017}. We verified the low level of vibrations with a fiber-based optical interferometer (attocube, FPS3010) that measures the displacement of the cold plate with respect to the optical table. The resulting time trace at the measurement bandwidth of $100$~kHz is presented in Fig.~\ref{fig2}(a). The data reveal a $1$~Hz periodic pattern of mechanical pulses characteristic of closed-cycle coolers. These pulses excite a set of high frequency vibrations resulting in a peak-to-peak (p-p) amplitude below $10$~nm at the given full bandwidth. This amplitude quickly rings down leading to a rather inhomogeneous distribution of vibrations in time with rms displacement fluctuations of $2.2$~nm. Although this is not entirely surprising, it is worthwhile to point out that the displacement fluctuations (often referred to as \textit{vibrational noise}) arise from a set of mechanical oscillators being excited by an impulse force during the cryo-cooler cycle and does not follow a well-defined statistical distribution, such as the Brownian noise.

The mechanical vibrations become naturally exacerbated when mechanical degrees of freedom are added to the system. In our case, the addition of nanopositioners is necessary to provide in-situ control. As a demonstration, we measure the displacement amplitude along the z-axis of a xyz-nanopositioner set (attocube, two ANPx101 and ANPz102) directly mounted on the cold plate. The time trace in Fig.~\ref{fig2}(b) shows that the p-p displacement increases significantly by more than a factor of $3$, reaching up to a maximum of around $35$~nm at each pulse of the cryo-cooler. We note that in contrast to the p-p amplitude at the mechanical pulses, the displacement amplitude between two pulses remains fairly similar, with an increase in rms fluctuations by less than factor of two to a value of $3.4$~nm.
These results make clear that additional measures are needed for the cavity setup, shown in Fig.~\ref{fig1}(b), to be operational in the cryostat. Without any modifications, the rigid fiber mirror stack would closely follow the motion of the cold plate and thus feature fluctuations as in Fig.~\ref{fig2}(a), whereas the planar mirror stack would vibrate similar to the nanopositioner stack with characteristics in Fig.~\ref{fig2}(b). The differential motion between the two cavity mirrors is expected to be of the order of tens of nm, rendering experiments even with a low finesse cavity impossible.

To improve the mechanical stability, we designed and implemented a spring based vibration isolation stage that isolates the cavity setup from vibrations of the cold plate. A sketch including the cavity setup and the spring stage is depicted in Fig.~\ref{fig1}(c). The cavity setup of Fig.~\ref{fig1}(b) is represented here by two springs, corresponding to the fiber mirror stack and the planar mirror stack with their respective spring constants k$_{F}$ and k$_{M}$ (where rigid fiber mirror stack implies $\mathrm{k}_{F}\gg\mathrm{k}_{M}$). This full cavity setup is mounted on a titanium plate, labeled as a spring table in Fig.~\ref{fig1}(c). The spring table rests on a set of four springs (represented by a single spring in the sketch), each of which has a nominal spring constant of k$_{S} $ $= 1.52$~N/mm at ambient conditions. To suppress high oscillation amplitudes at the spring stage resonance, we include a soft magnetic eddy-current damper, denoted with damping constant $\gamma$. The total mass compressing the springs is $0.51$~kg, yielding a nominal resonance frequency of the vibration isolation stage of $18$~Hz. We choose the payload and spring constant such that the resulting resonance frequency lies sufficiently far above the resonance frequency of the optical table ($5$~Hz) to suppress their coupling. The spring system acts as a mechanical low pass filter, that suppresses vibrations by a factor $100$ per decade above resonance frequency with a transition to suppression by a factor $10$ per decade towards higher frequencies due to the eddy-current damping. This means that contributions to the mechanical vibrations will become suppressed from tens of nm down to the order of $0.1$~nm for the frequencies above $180$~Hz.

An important challenge is to ensure reliable thermalization of the cavity with the cryostat cold plate without reducing the vibration isolation by spring stage. This is crucial because an efficient thermal link inevitably constitutes an additional mechanical connection than can potentially allow vibrations to bypass the vibration isolation stage. We address this challenge by attaching, to the spring table, a bundle of thin copper lamella with high thermal conductivity that are sufficiently soft and flexible to suppress transfer of mechanical vibrations.

To characterize the performance of the complete assembly, we measured the relative displacement between the fiber and macroscopic mirrors using the cavity itself as an interferometer. A schematic overview of the experiment on the optical table is presented in Fig.~\ref{fig1}(c). The optical excitation is provided through the fiber side of the cavity. The transmitted light through the planar mirror of the cavity is guided to a photodiode (Siemens, BPW34 with DL Instruments, 1211 Current Preamplifier) and/or is spectrally dispersed by a monochromator (Roper Scientific, Acton SpectraPro-275) and detected by a CCD camera (Princeton Instruments, Spec-10). A part of the transmission signal can optionally be guided to another photodiode via a beamsplitter which can then be used to perform active feedback stabilization of the cavity length using the piezo actuator below the fiber mirror. 

For the stability characterization measurements, we used a fiber with a dielectric coating (LaserOptik, T $= 0.0032$) and a radius of curvature (ROC) of \SI{14}{\micro\meter}. The macroscopic mirror has a silver coating (T $\simeq 0.008$) which leads to higher photon leakage towards the free-space side. The laser light at $780$~nm (MSquared, SolsTiS) is coupled to the cavity through the fiber. A finesse $\mathcal{F}$ = 110 is achieved at the cavity length $L \simeq$ \SI{5}{\micro\meter}, corresponding to a mode number $q = 13$. For the stability characterization measurements, the transmission at the position of maximum slope of resonance is monitored as a function of time, then the fluctuations in the transmission signal are converted to cavity length fluctuations (see Appendix~\ref{conversion} for the conversion of the transmission signal to displacement).

Figure~\ref{fig2}(c) depicts the cavity length fluctuations as a function of time with measurement bandwidth of $100$~kHz. Compared to the measurements on a single set of positioners in Fig.~\ref{fig2}(b), the displacement is improved drastically such that it is almost invisible when compared at the same scale. In Fig.~\ref{fig2}(c), the inset on the right shows the time trace with a $20$-fold magnification. Here, it becomes visible that the vibration kicks at interval of $1$~s are strongly suppressed down to a p-p amplitude of $0.3$~nm. The rms cavity length fluctuations are $31.6 \pm 0.5$~pm at an integration bandwidth of $100$~kHz. The measurement bandwidth is large such that it encompasses all significant mechanical resonances of the system. However, the observed stability of the cavity can improve if the manipulation or measurement is performed at smaller bandwidth in a particular experiment. This is shown in Fig.~\ref{fig2}(d) where we present a direct comparison of the rms vibration level as a function of integration bandwidth for the cavity platform (circles), the set of nanopositioners (triangles) and the bare cold plate (squares). Here, we see that the implementation of the vibration isolation stage yields an improvement of almost two orders of magnitude across the entire bandwidth. 

We identify the source of the remaining vibrations by analyzing the Fourier transform (FT) in Fig.~\ref{fig2}(e) for the cavity length fluctuations shown in Fig.~\ref{fig2}(c). The peaks around $20$~Hz can be assigned to resonances of the vibration isolation stage. The stage excites both mirror stacks, and a small but visible relative displacement of the two mirrors is observed. Starting from $250$~Hz, more peaks are visible in the FT, eluding a detailed analysis. Some can presumably be attributed to the various components in the setup, but also the thermal links possibly provide a path for vibrations to be transmitted from the cold plate to the cavity that may not get fully filtered by the vibration isolation stage. Finally, there are also several sharp resonance peaks which are ascribed to the higher harmonics of $50$~Hz. 

\begin{figure}[t]
\centering
\includegraphics[scale=0.96]{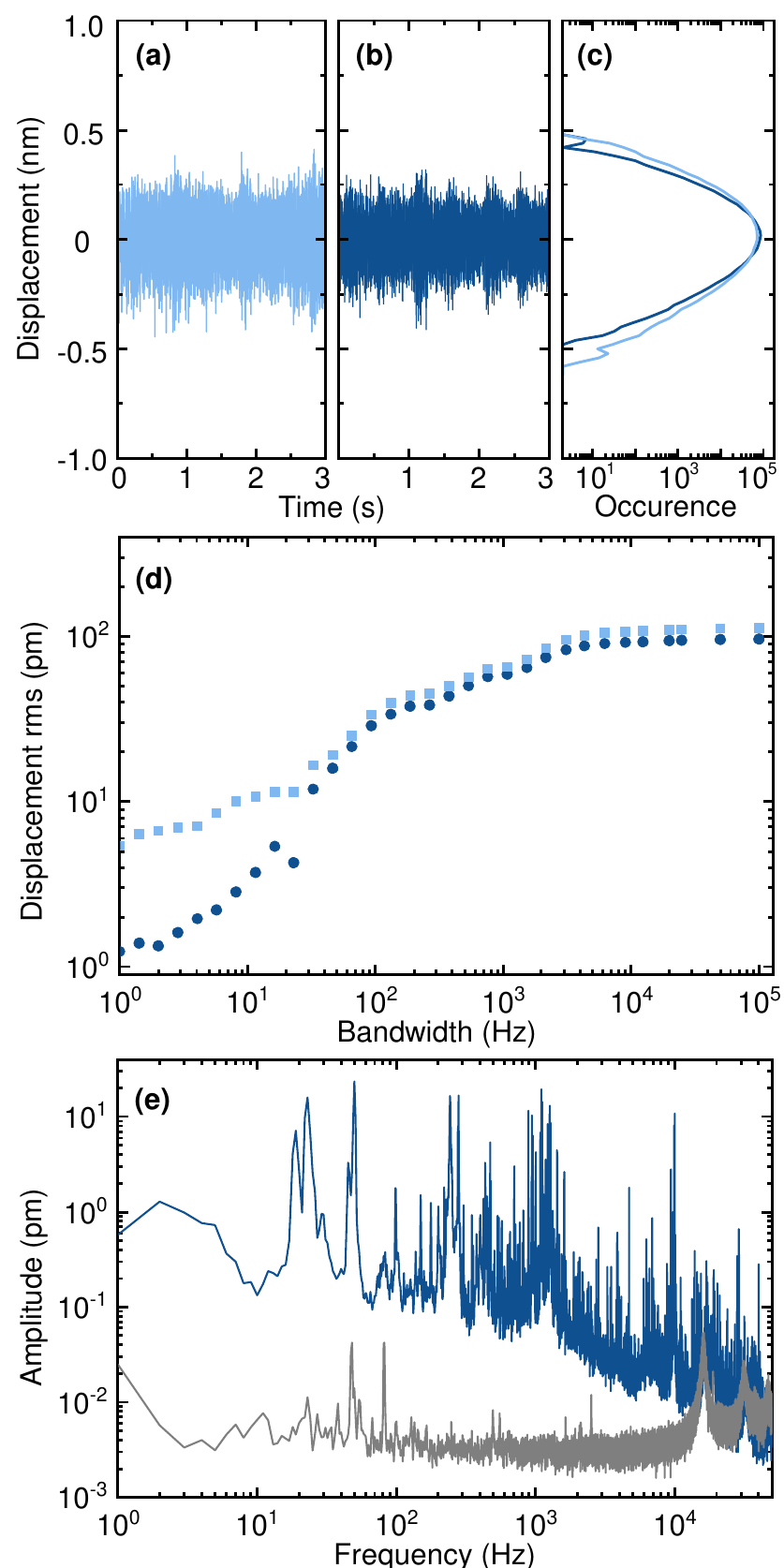}
\caption{Mechanical vibration characteristics at cryogenic temperature. The vibration displacement of the tunable cavity at $6.5$~K (a) without lock and (b) with lock at a bandwidth of $100$~kHz. (c) Occurrence of cavity length fluctuations over $10$~s. (d) Displacement rms of cavity fluctuations as a function of integration bandwidth without lock (square) and with lock (circle). (e) Fourier transform of cavity fluctuations with lock (blue) and noise floor (grey) with a frequency resolution of $1$~Hz.}
\label{fig3}
\end{figure}

We note that all preceding measurements with data in Fig.~\ref{fig2} were performed at room temperature by turning on the cryo-cooler for several minutes to evaluate the vibrations without performing a full cool-down to the base temperature. This is a useful tool for a quick characterization, especially during the development stage.  
In the following, we characterize the actual performance at cryogenic conditions where the mechanical properties of various components might differ with change in temperature and pressure, and result in different mechanical fluctuation characteristics. To this end, the cavity setup was cooled down to $6.5$~K in vacuum (the nominal base temperature of about $4$~K was not reached in this specific experiment due to a temporary modification of the radiation shield for a wire feed-through), with cryogenic characteristics shown in Fig.~\ref{fig3} for an integration bandwidth of $100$~kHz. Figure \ref{fig3}(a) shows the cavity length fluctuations without active stabilization with p-p amplitude and rms fluctuation values of $0.7$~nm and $117 \pm 7$~pm, respectively.

Subsequently, we evaluated the effect of the active damping where a part of the transmission signal was used to feedback the piezo actuator to stabilize the cavity length with a home-made servo. As shown in Fig.~\ref{fig3}(b), the length fluctuations were damped by active feedback. While the p-p amplitude remained at $0.7$~nm, the rms fluctuations reduced to $89 \pm 5$~pm, or by $25 \%$. The improvement can be visualized by counting the occurrences of each cavity displacement amplitude over a full time trace as  in Fig.~\ref{fig3}(c): The tail of the peak shape is curtailed, indicating an increased suppression with increasing fluctuation amplitudes. 

The comparison of the rms fluctuations as a function of bandwidth is shown in Fig.~\ref{fig3}(d). The active feedback stabilization works efficiently up to $\sim 50$~Hz, indicating an upper limit for the bandwidth of the feedback at that frequency. However, the analysis of rms fluctuations as a function of bandwidth shows that the mechanical stability of around $1$~pm can be reached with small measurement integration bandwidth. The FT of the cavity fluctuations with an active feedback is shown in Fig.~\ref{fig3}(e). Here, we observe similar features as in the room-temperature data of Fig.~\ref{fig2}(e), namely a relatively strong increase in rms fluctuations at low frequencies ($10 - 100$~Hz), where the resonance of the vibration isolation stage at around $20$~Hz and a sharp peak around $50$~Hz dominate. For frequencies above $200$~Hz, two sets of resonances dominate the vibrations, one of which lies in the range $\sim 200-500$~Hz and the other between $\sim 1-2$~kHz. In Fig.~\ref{fig3}(d), a step-wise increase in the displacement rms can be clearly observed at these frequency ranges. As discussed above, those contributions presumably arise from the mechanical resonance of individual components of the cavity setup and, potentially, from vibrations of the cryostat that get transferred via the thermal links. At $10$~kHz, the sharp peak originates from the carrier frequency of the pulse width modulator built in the inverter driving the $1$~Hz rotating valve of the three phase motor of the cryostat.

\section{APPLICATION: STRONG-COUPLING BETWEEN CAVITY AND SEMICONDUCTOR MONOLAYER}
To highlight the functionality of our cryogenic open-cavity we demonstrate the formation of exciton-polaritons in monolayer WSe$_2$ in the strong-coupling regime. Monolayer WSe$_2$ is representative for the class of atomically thin TMD materials with a direct band-gap \cite{Mak2010,Splendiani2010}, known for sizable light-matter interactions mediated by excitons \cite{Wang2018}. In addition to providing a light-matter interface, excitons in TMD monolayers exhibit unique valley properties at distinct points in momentum space, which can be selectively excited by circularly polarized light for opto-valleytronic applications \cite{Xiao2012,Xu2014}. Strong-coupling of the fundamental exciton transition to cavities have been demonstrated in various realizations  \cite{Schneider2018}, including monolithic photonic crystal \cite{Zhang2018}, dielectric \cite{Liu2015} and plasmonic \cite{Lundt2016,Cuadra2018} cavities at room-temperature as well as open-cavities both at room-temperature \cite{Flatten2016,Gebhardt2019} and under cryogenic conditions \cite{Dufferwiel2015,Sidler2017}. None of these experiments, however, combined the open-cavity concept with cryogenic operation in a closed-cycle cryostat.
 
The WSe$_2$ monolayer was exfoliated, encapsulated in hexagonal boron nitride (hBN) and transferred onto a silver mirror (see Appendix~\ref{sample} for details). The neutral exciton transition $X$ with energy $E_X$ = 1.725~eV and full-width at half-maximum linewidth $\Gamma = 6.1$~meV was identified with confocal cryogenic spectroscopy in differential reflection (see Appendix~\ref{coupling} for details). After confocal characterization, the silver mirror with the WSe$_2$ heterostack was inserted into the closed-cycle cryogenic cavity setup and paired with a fiber mirror with ROC of \SI{14}{\micro\meter} and silver-coating (T $\simeq 0.005$). The finesse of the cavity was measured to be $\mathcal{F} = 30$. 

\begin{figure}[t]
\centering
\includegraphics[scale=0.95]{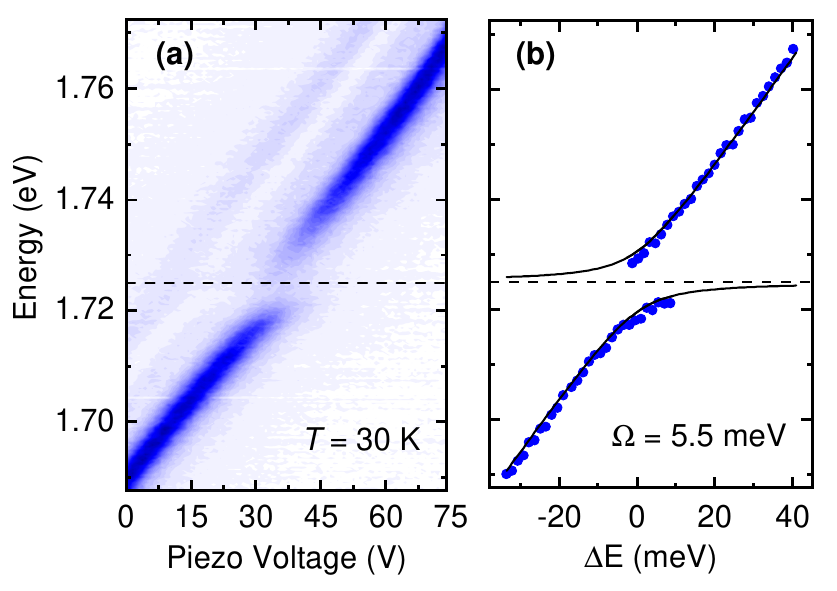}
\caption{Strong-coupling of monolayer WSe$_2$ in the closed-cycle cryogenic cavity. (a) Color-coded transmission through the coupled cavity-monolayer system as function of piezo voltage tuning of the photonic resonance energy. The avoided crossing as a characteristic signature of strong-coupling is observed on resonance with the exciton energy at $1.725$~eV (shown as dashed line). (b) Energy of the exciton-polariton peak extracted from the data in (a) as a function of energy detuning. The solid line corresponds to the coupled oscillator model with a normal-mode splitting of $\Omega = 5.5$~meV.}
\label{fig4}
\end{figure}

The cavity setup was subsequently cooled down to a temperature of $30$~K. We identified the monolayer region by performing a differential absorption scan in lateral dimension across the sample using white-light excitation (NKT, SuperContinuum Laser) through the fiber and recorded the spectrally-dispersed transmission signal. The cavity length for the scan was set to about \SI{50}{\micro\meter} such that it was not in the stable cavity regime. Once the monolayer was located inside the cavity, the cavity length and thus the effective cavity mode volume was reduced. At the cavity length of \SI{2.9}{\micro\meter} corresponding to the piezo voltage of $\sim 15$~V in Fig.~\ref{fig4}(a), the cavity resonance was identified at $E_C \simeq 1.7$~eV with a linewidth of $\kappa$ = 6.3~meV limited by the cavity finesse of $30$. 

We monitored the cavity resonance mode as it was tuned through the exciton resonance with the piezo voltage, the corresponding transmission is shown in Fig.~\ref{fig4}(a) for the cavity mode $q = 8$. The transmission at large energy detuning between the photon and exciton energy $\Delta = E_C -E_X$, i.e. the piezo voltage of $< 30$~V or $> 60$~V, represents the uncoupled photonic resonances inside the cavity with the brightest peak corresponding to the fundamental transverse electromagnetic mode accompanied by two weak, blue-shifted higher order modes. At small energy detuning, the normal-mode splitting emerges as a hallmark of the strong-coupling regime and exciton-polariton formation. The peak energy of the coupled system were extracted and plotted in Fig.~\ref{fig4}(b) as a function of resonance detuning. The solid red lines represent the best fit of the coupled oscillator model to the data with a Rabi splitting of $\Omega = 5.5$~meV (see Appendix~\ref{coupling} for model details). Using the photonic and excitonic linewidths of $\kappa = 6.3$ and $\Gamma = 6.1$~meV, respectively, the corresponding cooperativity of the system calculates to $C = 2 \Omega^2/(\kappa \Gamma) = 1.6$ and establishes the strong-coupling regime of exciton-photon interaction.

For the observation of the strong-coupling limit, both the open-cavity in-situ tuning and the length stability were crucial. As the first key feature, our platform provides an effective fine-tuning range of the cavity length over $4.8$~mm with a combination of nanopositioner and piezo element as an enabling factor for identifying relevant monolayer regions at long cavity lengths and subsequently observing strong-coupling with precision tuning of the spectral resonance condition. This is highlighted in Fig.~\ref{fig4}(a) via piezo-controlled tuning of the cavity length by $\sim 120$~nm, which corresponds to a cavity resonance detuning from $700$ to $733$~nm for mode $q = 8$. Second, the cavity stability is also critical for strong-coupling. For example, without vibration isolation as in Fig.~\ref{fig2}(b), the mechanical vibrations of $30$~nm would exceed the cavity spatial-equivalent linewidth of $13$~nm (for a finesse of $30$) and thus effectively reduce the coherent coupling. Moreover, the cavity length tuning range of $\sim 120$~nm in Fig.~\ref{fig4}(a) would only be resolved up to roughly four steps, rendering a controlled detuning of the spectral resonance condition close to impractical.

\section{CONCLUSIONS}
In conclusion, we have reported the layout and operation of a fully tunable fiber-based Fabry-P\'{e}rot cavity in a closed-cycle cryostat as a platform for quantum optical technologies and cavity QED experiments. The platform provides a spatial tuning range of $6$~mm over the sample surface and $4.8$~mm between the two mirrors as well as additional tuning capability of the aspheric lens for an efficient coupling of light into free-space through the planar mirror. With the combination of passive and active vibration isolation techniques, we achieve deterministic and reproducible mechanical stability in form of p-p amplitude of $0.7$~nm and rms fluctuations of $89$~pm in the bandwidth of $100$~kHz over a complete cycle of the closed-cycled cryostat at $6.5$~K. This level of mechanical stability in a high measurement bandwidth corresponds to an upper limit of the cavity finesse in the order of $10^4$ for a signal-to-noise ratio of $1$, while maintaining sample positioning in all three spatial degrees of freedom over several mm. These stability numbers quality our platform for state-of-the-art cavity QED experiments with Q-factors in the range of $10^5 - 10^6$ and in-situ tunability for a wide variety of solid-state emitters such as quantum dots and wells in conventional semiconductors, TMDs, color centers or rare-earth ions \cite{Barbour2011, Besga2015, Riedel2017, Foerg2019, Bogdanovic2017, Najer2019, Tomm2021, Merkel2020, Salz2020, Casabone2020, Ruf2021}. To highlight this controlled interaction, we demonstrated the application of our platform to exciton-polariton formation in the specific case of monolayer WSe$_2$.

With these results, we present our versatile open-cavity platform as a viable tool to study light-matter interactions at low temperatures. It can accommodate more complex materials such as van der Waals heterostructures \cite{Foerg2019,Ashida2020,Kennes2021} with unique nonlinearities \cite{Zhang2021} and is ideally suited to advance quantum applications based on single-photon sources \cite{Lodahl2015}, novel qubits of rare-earth ions \cite{Wesenberg2007,DeRiedmatten2008,Kolesov2012,Kindem2020} or respective light-matter interfaces for quantum networks \citep{Reiserer2015,Daiss2021}. Future improvements in mechanical vibration management by advanced passive isolation strategies and optimized active feedback control will place our open-cavity platform in a closed-cycle cryostat at the forefront of enabling technologies for quantum science and related applications. 

\section{Acknowledgments}
This research was funded by the European Research Council (ERC) under the Grant Agreement No.~772195, as well as the Deutsche Forschungsgemeinschaft (DFG, German Research Foundation) within the Priority Programme SPP 2244 ``2DMP'' and the Germany's Excellence Strategy EXC-2111-390814868. S.\,V. has received funding from the European Union's Horizon 2020 research and innovation programme under the Marie Skłodowska-Curie grant agreement Spin-NANO No 676108. H.\,T., K.\,K. and D.\,H. acknowledges funding from the European Union's FET Flagship on Quantum Technologies with Grant Agreement No. 820391 (SQUARE). D.\,H. acknowledge financial support by the Bundesministerium f\"ur Bildung und Forschung via Q.Link.X. K.\,W. and T.\,T. acknowledge support from the Elemental Strategy Initiative conducted by the MEXT, Japan, Grant Number JPMXP0112101001, JSPS KAKENHI Grant Numbers JP20H00354 and the CREST(JPMJCR15F3), JST.

\appendix
\vspace{-11pt}
\section{Mirror and sample details} \label{sample}
\vspace{-11pt}
The fiber micro-mirrors were laser-profiled to concave shape with a CO$_2$ laser. Fiber mirror for stability measurement had DBR coating from LaserOptik. Fiber mirror for strong-coupling experiment had a $60$~nm silver layer deposited using evaporation. A thin SiO$_{2}$ layer was evaporated on top of fiber micro-mirror for protection. The macroscopic mirror was formed by a $55$~nm thick silver layer evaporated onto a glass substrate (Laseroptik) and coated with $125$~nm of SiO$2$. The fiber micro-mirror in the strong-coupling experiment had a slightly thicker silver layer compared to the planar mirror to ensure dominant cavity out-coupling through the latter one. The WSe$_2$ monolayer (HQ graphene) and hBN (NIMS) for encapsulation were exfoliated with the scotch tape method and subsequently assembled into a heterostack with the hot pick-up technique \cite{Pizzocchero2016}. The heterostack was transferred onto the mirror and sequentially annealed in vacuum at $200^{\circ}$~C for $12$ hours. 

\vspace{-11pt}
\section{Displacement analysis via cavity interferometry} \label{conversion}
\vspace{-11pt}
The fluctuations of the cavity length due to external mechanical disturbances can be measured by using the cavity as a displacement interferometer. To determine the cavity displacement induced by mechanical vibrations, the transmission signal through the planar mirror is recorded at the slope of a cavity resonance where the sensitivity to changes in transmission as a function of the cavity length is highest. Changes in transmission are subsequently converted to changes in the cavity length.

\begin{figure}[b!]
\centering
\includegraphics[scale=1]{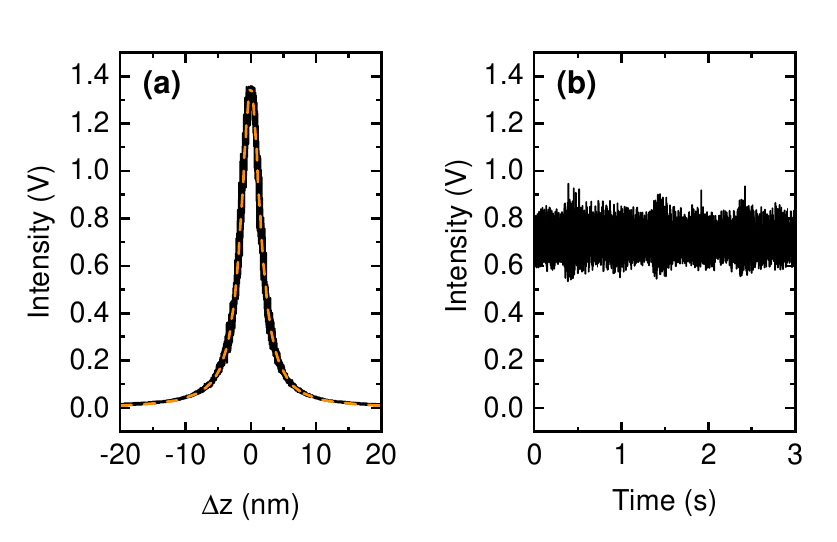}
\caption{(a) Transmission signal as a function of cavity length. The solid black line is the data around a cavity resonance obtained by applying a voltage to the piezo actuator. The orange dashed line shows the corresponding fit to the Fabry-P\'{e}rot transmission function of Eq.~\ref{form1} with $\mathcal{F}=110$. (b) Transmission signal measured with active stabilization at the maximum slope of the cavity resonance shown in (a) at a temperature of $6.5$~K. The measurement bandwidth was $100$~kHz.}
\label{SIfig1}
\end{figure}

To this end, the transmission signal across a cavity resonance is recorded while sweeping the voltage applied to the piezo below the fiber that acts as an actuator. The cavity resonances are separated by the free spectral range (FSR) of $\lambda /2$. The finesse of the cavity resonance is determined from the transmission signal as a function of the cavity length as \cite{Saleh2019}:
\begin{equation}
T(z) = \frac{T_{0}}{1+\left(G\sin \Phi\right)^{2}},
\label{form1}
\end{equation} 
\noindent where $T _{0} $ corresponds to the transmission on resonance, $\Phi (z) = 2\pi (z-L)/\lambda$ depends on $z$ via the on-resonance cavity length $L$ and the wavelength $\lambda$, and $G$ is related to the finesse as $G = 2 \mathcal{F} /\pi $. The derivative of the transmission function (Eq.~\ref{form1}) is given by:
\begin{equation}
\frac{dT}{dz}=-\frac{4G^{2}\pi}{\lambda}\frac{\sin(\Phi)\cos(\Phi)}{\left[ 1+\left( G\sin \Phi\right)^{2} \right]^{2}}.
\end{equation}
\noindent Any change in the transmission as a function of time is converted to a change in the cavity length using the above derivative of the transmission function in a linear approximation of small fluctuations around the half-maximum position. 

\begin{figure}[t!]
\centering
\includegraphics[scale=0.95]{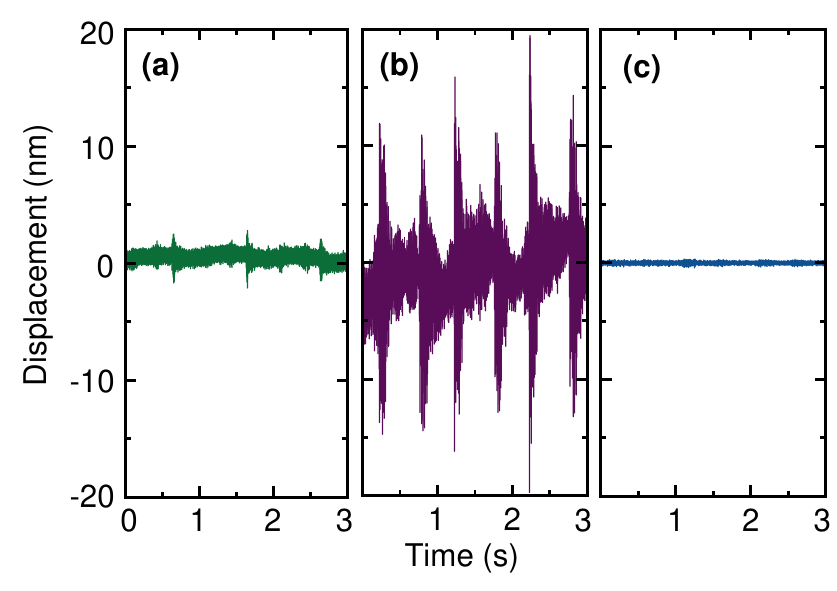}
\caption{Mechanical vibration characteristics at cryogenic temperature for (a) the cold plate, (b) a stack of nanopositioners on the cold plate, and (c) the cavity length fluctuations of the tunable open-cavity setup [same data as in Fig.~\ref{fig3}(b)].}
\label{SIfig2}
\end{figure}

In our experiments, the transmission signal was measured with a time resolution of \SI{10}{\micro\second}. The rms displacement fluctuations were calculated from the cavity length fluctuations as a function of time as:
\begin{equation}
d_{rms} = \sqrt{\frac{1}{n} \sum_{i=1}^{n} d_{i}^{2}},
\end{equation}
\noindent where $n$ is the total number of data points acquired over $10$~s and $d_i$ is the cavity displacement at each point. It is important to note that the cavity length fluctuations should be measured with an appropriate dynamic range defined by the finesse of the cavity. For example, the measurements shown in this work were performed with $\mathcal{F} = 110$ leading to the spatial-equivalent cavity linewidth of $\Delta L \simeq 3.5$~nm. Thus, the maximum displacement of the cavity length for a vibration kick of $0.7$~nm is still well within the interferometric measurement range of the cavity as illustrated in Fig.~\ref{SIfig1}(b). For completeness, Fig.~\ref{SIfig2} shows low-temperature mechanical vibration characteristics of the three configurations discussed in Fig.~\ref{fig2}. The p-p displacement for the cold plate [Fig.~\ref{SIfig2}(a)], a stack of nanopositioners [Fig.~\ref{SIfig2}(b)] and the open-cavity with vibration isolation [Fig.~\ref{SIfig2}(c)] is similar to room temperature, and the improvement in displacement fluctuations due to vibration isolation holds for both temperature ranges.

\begin{figure}[t!]
\centering
\includegraphics[scale=1]{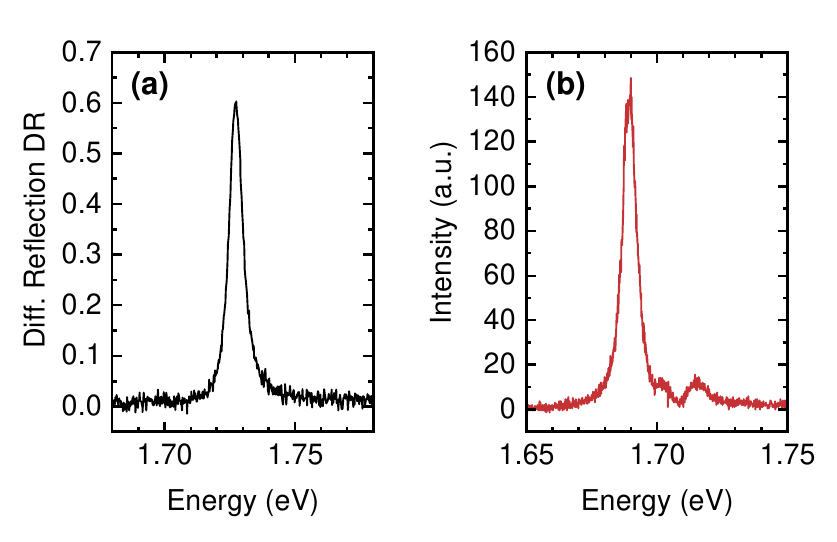}
\caption{(a) Differential reflection (DR) spectrum of monolayer WSe$_{2}$ recorded with cryogenic confocal spectroscopy $4$~K. (b) Transmission spectrum of the cavity at $6.5$~K red-detuned from the exciton resonance.}
\label{SIfig3}
\end{figure}

\begin{figure}[b!]
\centering
\includegraphics[scale=0.93]{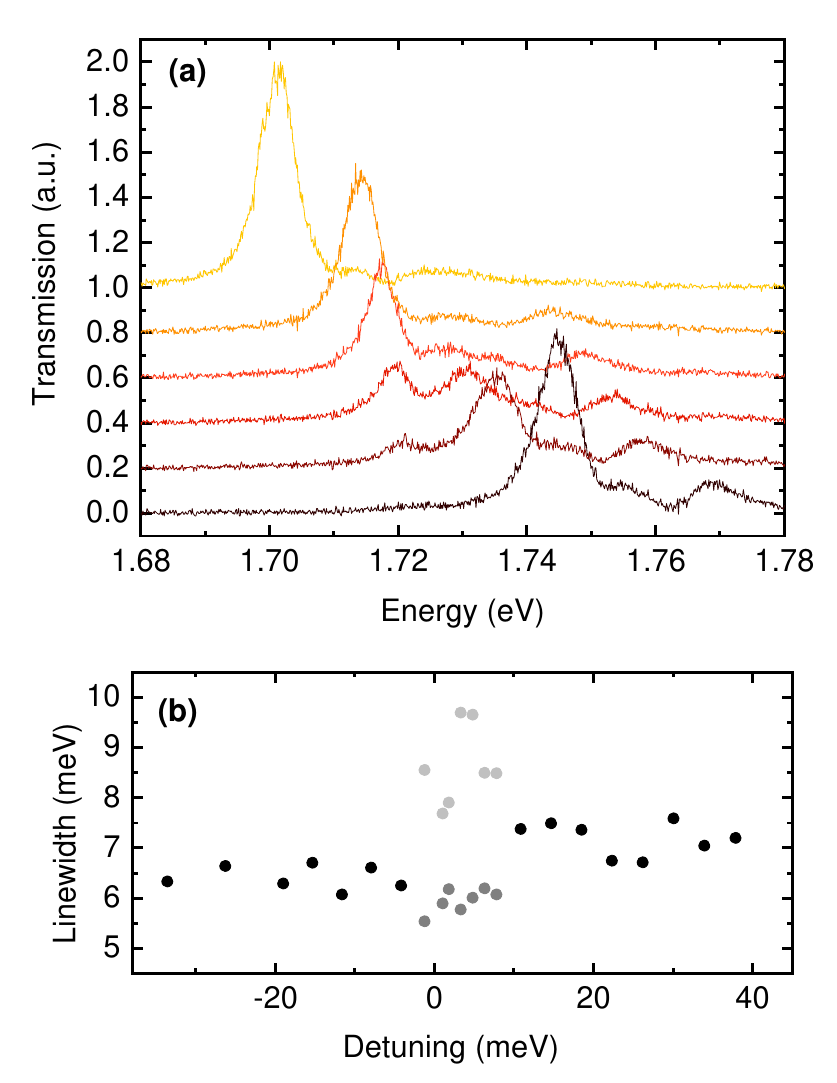}
\caption{(a) Representative transmission spectra at different exciton-cavity detunings around the resonance. (b) Linewidth of the upper (light grey) and lower (grey) polariton branches extracted from transmission spectra as a function of resonance detuning.}
\label{SIfig4}
\end{figure}
\vspace{-11pt}
\section{Characterization of exciton-cavity coupling} \label{coupling}
\vspace{-11pt}
The characterization of the fundamental exciton transition in monolayer WSe$_2$ was performed in a helium bath cryostat at $4$~K with a home-built confocal microscope. For reflection spectroscopy a halogen lamp or a supercontinuum laser (NKT, Supercontinuum Extreme) were used as broadband excitation sources with an illumination spot diameter of $\sim 1~\mu$m (attocube, LT-APO objective with numerical aperture of $0.82$). The reflected light was spectrally dispersed and detected by a nitrogen cooled-CCD (Roper Scientific, Acton SP2500 with Spec 10:100BR/LN). A typical differential reflection spectrum of monolayer WSe$_2$ is shown in Fig.\ref{SIfig3}, with the single resonance at $1.725$~eV and full-width at half-maximum linewidth $\Gamma  = 6.1$~meV corresponding to the neutral exciton $X$ transition. Figure \ref{SIfig3}(b) shows the primary cavity resonance at $E_C = 1.690$~eV (red-detuned from $E_X$) with a linewidth of $\kappa = 6.3$~meV as well as two weak, blue-shifted higher-order modes.

The exciton-photon coupling was modeled using a coupled oscillator model. The Hamiltonian of a two-level exciton interacting with quantized photonic mode in an optical cavity is the well-known Jaynes-Cummings Hamiltonian of an interacting system:
\begin{equation}
H= E_{C} (a^{\dagger} a) + E_{X} (x^{\dagger} x) + \hbar g \left(a^{\dagger} x+ a x^{\dagger}\right),
\end{equation} 
\noindent where $E_C$ ($E_X$) is energy of the cavity mode (exciton), $a$ and $a^{\dagger}$ ($x$ and $x^{\dagger}$) are the photon (exciton) annihilator and creation operators, respectively, and $g$ is the interaction strength. The solution to the above Hamiltonian in the presence of exciton and cavity decay rates $\Gamma$ and $\kappa$ (both in units of energy) leads to two new eigenstates of the coupled system referred to as the upper and lower polaritons with eigenenergies \cite{Kavokin2017}:
\begin{equation}
E_{\pm}=\frac{E_{C}+E_{X}}{2}-\frac{i}{2}\left(\kappa+\Gamma\right) \pm \Omega,
\label{PolaritonEnergy}
\end{equation}
\noindent where $\Omega$ is the complex Rabi splitting (in units of energy) given by:
\begin{equation}
\Omega = \sqrt{\textrm{g}^{2}+\left[\Delta -\frac{i}{2} \left( \kappa- \Gamma\right) \right]^{2}}.
\label{RabiFrequency}
\end{equation}
\noindent Here, $\Delta = E_C -E_X$ is the energy detuning from the cavity-exciton resonance condition. Coincidentally, the exciton and cavity decays were almost the same in our experiment, leading to a vanishingly small imaginary part for $E_{\pm}$ and $\Omega$. Using $\kappa=6.3$ and $\Gamma=6.1$~meV, best fit to the coupled system with data in Fig.~\ref{fig4} was obtained with Eq.~\ref{PolaritonEnergy} for a Rabi splitting $\Omega = 5.5$~meV at zero detuning $\Delta$. The transmission spectra and polariton linewidth analysis are shown in Fig.~\ref{SIfig4}. Note that the upper polariton linewidth is masked by higher order cavity modes.

%

\end{document}